\documentclass[aps,amsmath,twocolumn]{revtex4}
\usepackage{graphicx}
\begin{document}

\title{Hole-burning in an Autler-Townes doublet and in superluminal (subluminal) Electromagnetically induced transparency of a light pulse via a joint nonlinear
coherent Kerr effect and Doppler broadening}
\author{Bakht Amin Bacha}
\affiliation{Department of Physics, Hazara University, Pakistan}
\author{Fazal Ghafoor}
\affiliation{Department of Physics, COMSATS Institute of
Information Technology, Islamabad, Pakistan}
\author{Iftikhar Ahmad}
\affiliation{Department of Physics, Malakand University, Chakdara
Dir(L) , Pakistan}
\begin{abstract}
We investigate the behavior of light pulse propagation in a
4-level double Lambda atomic system under condition of
electromagnetically induced transparency. The Fano type
interference effect and spectral hole burning appears in the the
dynamics of the absorption-dispersion spectra caused by the joint
nonlinear coherence Kerr effect and Doppler broadening. The
coherent Kerr effect exhibits an enhancement (reduction) in
superluminal (subluminal) in negative (in positive) group index
while the Doppler broadening generates multiple hole burning in
the Autler-Townes like spectra of this system. The hole burning in
addition with coherent Kerr effect on the spectral profile
influences the dynamics of subluminal and superluminal of the
probe pulse through the medium. The characteristics of
superluminality and subluminality modified by considering
cold-Kerr-free medium and hot-Kerr-dependent mediums. The light
pulse delays and advances in different regions of dispersion
medium with the Doppler broadening and coherent Kerr effect.
Consequently, the pulse delays by $49\mu s$, while advance by
$-91\mu s$, for a same set of parameters [note: a revised version
is under preparation]

\end{abstract}

\maketitle
\section{Introduction}
The amplitude and the phase control of the group velocity of light
in optical media have attracted  a lot of attention in the recent
years.
\begin{figure}[t]
\centering
\includegraphics[width=2.5in]{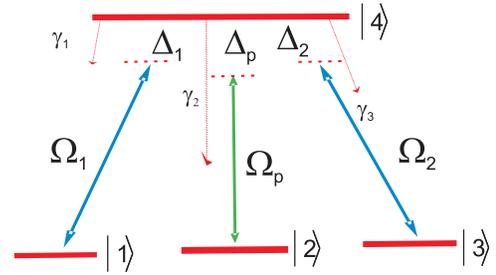}
\caption{(a) Schematics of the atomic system. (b)
Doppler-broadened system} \label{figure1}
\end{figure}
High degree control over this speed is possible and it can be made
much smaller than c, greater than c, or even negative
\cite{wb2009}. The manipulation of light pulse propagation in an
optical medium can be realized by changing the the dispersive
properties of the medium. The exact control over the optical
properties of the medium gives rise to the observation of some
interesting phenomena based on quantum coherence and quantum
interference\cite{EG1976,GWY1999}. Examples of some fascinating
phenomena, are Coherent population Trapping (CPT,
\cite{EG1976,GG1976} Lasing without Inversion (LWI)\cite{
se1989,AS1995,YF1996} Electromagnetically Induced Transparency
(EIT)\cite{ MO1994,SE1997,MF2005,MX19955} Multi-wave mixing \cite{
YJ2003,YA2008,LG2006} Enhancing Kerr nonlinearity\cite{YG2006}
Optical Soliton \cite{YW2005,GP2005}. The development of
theoretical and experimental techniques for control the light
pulse propagation through optical media are the results of  fast
few decades. Electromagnetic fields are used to create large
atomic coherence and it is possible to tailor the amplification,
absorption, dispersion properties of multilevel
atoms\cite{rw2002,se1997}. The intrusting application of these
techniques is to adjust the group velocity of light pulses to
propagate it very slowly or very fast. The region of normal
dispersion is the one in which group velocity is lesser than the
vaccum velocity of light $(v_g<c)$, while the region of anomalous
dispersion is the one in which $v_g>c$. In the normal dispersion
region subluminal and in the anomalous dispersion region
superluminal propigation of light occurs \cite{gs2004}. The well
 known approach to slow down the light pulse propagation in atomic
 vapor is the technique of electromagnetically induced
 transparency. Electromagnetically induced
transparency(EIT) and Spontaneously generated coherence(SGC)
change the steady state response of the medium\cite{cp2004}.
Experimentally Hau,\textit{et al} slow down the group velocity of
light pulse to $17ms^{-1}$ in bose Einstein condensate. In Rb
vapor, the group velocity of light was reduced to $( 90ms^{-1} ,
8ms^{-1})$\cite{AK95,OS96,LV99,MM99,DD99}. Agarwal \textit{et al}
received a group index of
 $10^3$ in two level atomic system in Doppler broadened configuration
of saturated absorption spectroscopy\cite{gs03}. Shang-qi kuang
\cite{shang08} theoretically present slow light propagation on
coherent hole burning in a Doppler broadened three level A-type
atomic configuration. They use the coherent hole burning dip and
achieved the slow light propagation at resonance condition.
 A large numbers
of experiments on slow light propagation in Doppler broadened
medium are reported by M.M Kash and A .kasapi\cite{mm99,ak95}. It
is pointed out that the Kerr nonlinearities could be increased by
several orders of magnitude by taking advantage of the EIT. Kang
and Zhu \cite{hkang2003} proposed a large enhancement in the Kerr
nonlinearity with vanishing linear susceptibility in coherent
prepared four-level Rb, atoms. Quantum coherence and interference
manifested by EIT, suppresses the linear
 susceptibility and greatly enhances the nonlinear susceptibility
 at low light intensity. Large kerr nonlinearity has also been reported
 using EIT, in \cite{yfchin2006}.
 Wang \textit{et al}. proposed \cite{zbwang2006} the use of double
 EIT schemes for the optimal production of the cross phase modulation. G.S Agarwall\cite{tdgs2007}
 reduced the group velocity
of light pulse by Kerr effect under the condition of EIT and found
that the group velocity considerably reduces in a dispersive
medium .Kocharovskaya \cite{ok99} stopped the light pulse in hot
gases. In a another experiment light has been slow down to
$57ms^{-1}$. Experimentally light has been stopped and stored in
atomic coherence\cite{ms03,LS01}. In a classic paper
\cite{Schmidt199}. Slow light has many potential application in
optical delay lines, quantum entanglement of slow photon, non
classical squeeze state,entangled atomic ensembles, optical black
hole\cite{LV99,ML00,MD00,UL02}. L.J Wang \textit{et
al.}\cite{Wang} demonstrated superluminal light propagation for
light using the region of lossless anomalous dispersion  between
two closely spaced gain lines in a double-peaked Raman gain
medium. Saharai \textit{et al}\cite{ Saharai} proposed a scheme
based on four-level EIT atomic medium and controlled the
normal/anomalous dispersion of light via phase of the driving
field. Theoretically four level gain atomic system is extend to
polychromatic pump fields and multiple anomalous region of
superluminality were observed in the gain doublet region as well
as between the two pairs of the doublet regions\cite{bab2013}.
Moti fridman \textit{et al} used the superluminal group velocity
for temporal cloaking\cite{Moti2012} follow the spatial cloaking
\cite{peny2006,Leon2006} idea in the temporal domain. A hole is
created space and time windows to hide the object and information
by the manipulation of positive negative group index. For best
cloaking, it is require to increase the time gap to microsecond
and to millisecond\cite{Robret2012}. The manipulation of negative
group velocity is also used for the quality of imaging. However in
their experiment there is also a lack of best quality of the
images arising due to low values of the negative group index as
discussed in thier experimental paper. According to these facts it
is necessary to manipulate  slow  and fast light for better
communication  and  better technology of imaging in cloaking.
\cite{Leonh2009,Fridman2012,Mc2011}. In this article we explore
the mechanism for slow and fast manipulation by the jointly effect
of kerr nonlinearity and Doppler broadening in a double lamda
atomic configuration. We show that how the slow and fast light can
be significantly influence by the jointly effect of kerr
nonlinearity and Doppler broadening.

 A question can be raised
that, is there a way to enhance the group velocity of a fast laser
light in a dispersive medium.If the answer is yes then, what
phenomenon is responsible for the increase in the group
velocity.The answer of this interesting question is explored in
the present article.In order to answer this question, we propose a
five-level atomic scheme,and the effect of Kerr nonlinearity on a
fast light is studied.The scheme is based on EIT in which the
group index can be enhanced significantly in the negative domain
due to the Kerr effect.This enhancement boosts superluminality in
a dispersive medium.
\section{Model and equation}
We consider experimental 4-level double lamda type
atomic-configuration driven by two appropriate coherent control
fields and a probe field [see Fig. 1]. The lower ground levels
$\left\vert 1\right\rangle $ and $\left\vert 3\right\rangle $ are
coupled with the upper excited level $ \left\vert 4\right\rangle $
by two control fields of Rabbi frequencies $\Omega _{1}$ and
$\Omega _{2}$, whereas the lower level $\left\vert 2\right\rangle
$ is coupled with level $ \left\vert 4\right\rangle $ by a probe
field of Rabbi frequency $\Omega _{p}$. To observe the condition
of atomic motion relative to the frequencies of the driving
fields, we modified
 the system interaction due to flexible environment.
Therefore we consider the atomic velocity linear in the response
function of the medium of the system respectively. To explain the
equations of motion and optical properties of the system, we
proceed with the following interaction picture Hamiltonian in the
dipole and rotating wave approximations:
\begin{eqnarray}
H(t)& =&-\frac{\hbar }{2}\Omega _{1}\exp[-i\Delta_1t] \left\vert
1\right\rangle \left\langle 4\right\vert \nonumber\\&&-\frac{\hbar
}{2}\Omega _{2}\exp[-i\Delta_2t ]\left\vert 3\right\rangle
\left\langle 4\right\vert\nonumber\\&& -\frac{\hbar }{2}\Omega
_{p}[-i\Delta_p t]\left\vert 2\right\rangle \left\langle
4\right\vert +H.c.
\end{eqnarray}%
The angular frequencies three fields are related to the atomic
states frequencies as: $\Delta_1=\omega_{14}-\omega_1$,
$\Delta_2=\omega_{34}-\omega_2$ and
$\Delta_p=\omega_{24}-\omega_p$. The general form of density
matrix equation is given by the following relation:
\begin{equation}
\frac{d\\\rho_t}{dt}=\frac{-i}{\hbar}[H_t
,\rho_t]-\frac{1}{2}\Gamma_{ij} \sum(  \sigma^\dagger  \sigma
\rho+\rho \sigma^\dagger \sigma-2\sigma \rho  \sigma^\dagger)
\end{equation}
where $\sigma^\dagger$ is raising operator $\sigma$, is levering
operator for the three decays.  After straight forward calculation
we obtain the three coupled rate equations in the first order as:
\begin{eqnarray}
\overset{\cdot }{\overset{\sim }{\rho
}}_{24}&=&[i\Delta_p+\frac{1}{2}(\gamma_1+\gamma_2+\gamma_3)]\widetilde{\rho}_{24}
-\frac{i}{2}\Omega_1\widetilde{\rho}_{21}\nonumber\\&&-\frac{i}{2}\Omega_2
\widetilde{\rho}_{23}+\frac{i}{2}\Omega_p(\widetilde{\rho}_{44} -
\widetilde{\rho}_{22}),
\end{eqnarray}

\begin{eqnarray}
\overset{\cdot }{\overset{\sim }{\rho
}}_{21}&=&[i(\Delta_p-\Delta_1)]\widetilde{\rho}_{21}
-\frac{i}{2}\Omega_1\widetilde{\rho}_{24}\nonumber\\&&+\frac{i}{2}\Omega_p
\widetilde{\rho}_{41},
\end{eqnarray}
\begin{eqnarray}
\overset{\cdot }{\overset{\sim }{\rho
}}_{23}&=&[i(\Delta_p-\Delta_2)]\widetilde{\rho}_{23}
-\frac{i}{2}\Omega_2\widetilde{\rho}_{24}\nonumber\\&&+\frac{i}{2}\Omega_p
\widetilde{\rho}_{43},
\end{eqnarray}
Taking $\Omega_p$, in the first order, while $\Omega_1$ and
$\Omega_2$, in all order of perturbation and consider the atoms
initially in the ground state $\left\vert 2\right\rangle$, while
the population initially in the other states are zero then the
following conditions are applicable to the density maxtrix.
$\widetilde{\rho}_{22}^{(0)}=1$, $\widetilde{\rho}_{44}^{(0)}=0$,
$\widetilde{\rho}_{41}^{(0)}=0$,  $\widetilde{\rho}_{43}^{(0)}=0$.
The above set of equations can be solved for
$\widetilde{\rho}_{24}^{(1)}$ using the relation.
\begin{equation}
Z(t)=\int^{t}_{-\infty}e^{-M(t-t^,)}Pdt^,=-M^{-1}Q,
\end{equation}
where $Z(t)$ and $Q$ are column matrices while Q is a 3x3 matrix.
The solution is written by:
\begin{eqnarray}
\widetilde{\rho}_{24}^{(1)}=[\frac{4i(\Delta_1-\Delta_p)(\Delta_p-\Delta_2)\Omega_p}{(\Delta_2-\Delta_p)
[4A_1(\Delta_p-\Delta_1)-i\Omega^2_1]+i(\Delta_p-\Delta_1)\Omega^2_2}]
\end{eqnarray}
To add the Doppler broadened effect in the atomic configuration,
we replace the detuning parameters by: $\Delta_1=\Delta_1+
\alpha_1 k_1 v$, $\Delta_2=\Delta_2+ \alpha_2 k_2 v$,
$\Delta_p=\Delta_p+ k v$, where,  $\alpha_{i=1,2}=1$, indicate
co-propagation direction of coherent fields to the probe field.
and counter propagating directions, while $\alpha_{i=1,2}=-1$,
show counter-propagation direction of coherent fields to the probe
field. Here $k_1$, $k_2$, $k_p$ are the wave vectors of the two
coherent fields and a probe field and for simplicity we put
$k_1=k_2=k_p=k$.
\begin{eqnarray}
\widetilde{\rho}_{24}^{(1)}(kv)=[\frac{-4Ti(\Delta_p-\Delta_2+k
v-\alpha_2kv)} {i\Omega^2_2T+A(\Delta_2-\Delta_p-k v+\alpha_2k
v)}]\Omega_p,
\end{eqnarray}
\begin{eqnarray}
A=-i\Omega^2_1+4B(\Delta_p-\Delta_1+k v-\alpha_1k v)
\end{eqnarray}
\begin{eqnarray}
A_1=\frac{2i\Delta_p+\gamma_1+\gamma_2+\gamma_3}{2}
\end{eqnarray}
\begin{eqnarray}
B=\frac{2i(\Delta_p+kv)+\gamma_1+\gamma_2+\gamma_3}{2}
\end{eqnarray}
\begin{eqnarray}
T=[k v-\Delta_1+\Delta_p-\alpha_1k v]
\end{eqnarray}

\section{susceptibility and group index}
The susceptibility is a response function of medium due to an
applied electric field. The susceptibility of our driven hot
atomic system is calculated to the first order in the probe field
fields and to the all order in the control field. The calculated
susceptibility for our hot atomic system is written as:
\begin{eqnarray}
\chi(kv)=\beta[\frac{-4Ti(\Delta_p-\Delta_2+k v-\alpha_2kv)}
{i\Omega^2_2T+A(\Delta_2-\Delta_p-k v+\alpha_2k v)}],
\end{eqnarray}
Where $\beta=\frac{2N|\mu_{24}|^2}{\epsilon_0\hbar}$  and $N$, is
the atomic number density of the medium and $\mu_{24}$, is dipole
moment between the level $\left\vert 2\right\rangle$, and
$\left\vert 4\right\rangle$. To introduce the effect of kerr field
in the system we expand $\chi_v$, with the intensity of the
control field $\Omega_1$ in the following passion in the
perturbation limit\cite{TNGS2007}.
\begin{eqnarray}
\chi^k(kv)=\chi^{(0)}(kv)+I\frac{\partial}{\partial
 I}\chi(kv)|_{I->0}:
\end{eqnarray}
 Where  $\chi^{(0)}(kv)$, is the probe equation of motion
without the kerr field field $I=|\Omega_1|^2$, intensity and
$I\frac{\partial^2\chi_{kv}}{\partial \Omega^2_1}$, is the term
contributed due to the kerr field. The Kerr field effected
susceptibility in the presence of the Doppler broadening is the
following:
\begin{eqnarray}
\chi^k(kv)=\beta[\frac{-4TiZ} {i\Omega^2_2T-4BTZ}
+\frac{-4\Omega^2_1TZ^2}{(i\Omega^2_2T -4ZTB)^2}]
\end{eqnarray}
\begin{equation}
Z=(kv-\Delta_2+\Delta_p-\alpha_2kv)
\end{equation}
When $v=0$, there is no Doppler broadening effect in the system.
The system is called cold  atomic system, if there is Doppler
broadening effect in the system, then it is called hot atomic
system. For cold atomic system, we represent susceptibility
without kerr effect is $\chi$, and in the presence of kerr effect
is $\chi^k$. These susceptibility are from the eq13 and eq15, when
one put $v=0$. The Doppler susceptibilities are the average of
$\chi_{kv}$, and $\chi^k_{kv}$, over the Maxwellian, distribution
and is describe bellow:
\begin{eqnarray}
\chi^{(d)}=\frac{1}{ V_D\sqrt{\pi }}\int^\infty_{-\infty} \chi(kv)
e^{-\frac{(kv)^2}{V^2_D}} d(kv)
\end{eqnarray}
\begin{eqnarray}
\chi^{(dk)}=\frac{1}{ V_D\sqrt{\pi }}\int^\infty_{-\infty}
\chi^k(kv) e^{-\frac{(kv)^2}{V^2_D}} d(kv)
\end{eqnarray}
 Where $V_D=\sqrt{K_BT\omega^2Mc^2}$, is the Doppler width. Where $\chi^{(d)}$ and $\chi^{(dk)}$ are the
 Doppler broadened susceptibilities without the kerr nonlinearity as well as in the presence of kerr nonlinearity.
\begin{equation}
N_g=1+2\pi
Re[\chi]+2\pi\omega_{24}Re[\frac{\partial\chi}{\partial\Delta}],
\end{equation}
\begin{equation}
 \tau_d=\frac{L(N_g-1)}{c}
 \end{equation}
 These are the mean results which will be analyzed and discussed in
details. $\tau_d$ is group delay/advance time. When its value is
positive it is called delay and if its value is negative it is
called advance time. To observe the nature of the pulse shape at
the output we used the transfer function. The output pulse
$S_{out}(\omega)$, after propagating through the medium can be
related to the input pulse $S_{in}(\omega)$ by the relation:
$S_{out}(\omega)=H(\omega)S_{in}(\omega)$. We choose a Gaussian
input pulse of the form:
\begin{eqnarray}
S_{in}(t)=\exp[-t^2/\tau^2_0]\exp[i(\omega_{24}+\xi)t],
\end{eqnarray}
where $\xi$, is the upshifted frequency of the empty cavity. The
Fourier transforms of this Gaussian function is then written by
$S_{in}(\Delta_p)=\frac{1}{\sqrt{2\pi}}\int^{\infty}_{-\infty}S_{in}(t)e^{i(\omega_{24}-\Delta_p)
t}d t$. The expression is reported after integration in the form
of probe detuning as:
\begin{eqnarray}
S_{in}(\Delta_p)&=&\tau _{0}/\sqrt{2}\exp [
-(\Delta^2_p+\xi^2-2\Delta_p\xi)\tau^2 _{o}/4]
\end{eqnarray}
Where $\Delta_p=\omega_{24}-\omega_p$. Using the convolution
theorem the output $S_{out}(t)$ can be written as:
\begin{eqnarray}
S_{out}(t)=\frac{1}{\sqrt{2\pi}}\int^{\infty}_{-\infty}S_{int}(\Delta_p)H(\Delta_p)e^{i(\omega_{24}-\Delta_p)
t}d\Delta_p
\end{eqnarray}
\begin{eqnarray}
H(\Delta_p)=\frac{T\exp[\frac{-i
K(\Delta_p)L}{c}]}{1-R\exp[\frac{-2i K(\Delta_p)L}{c}]}
\end{eqnarray}
Where $T$, $R$, are the transmission and reflection coefficients.
\section{Results and Discussion}
\begin{figure}[t]
\centering
\includegraphics[width=3.5in]{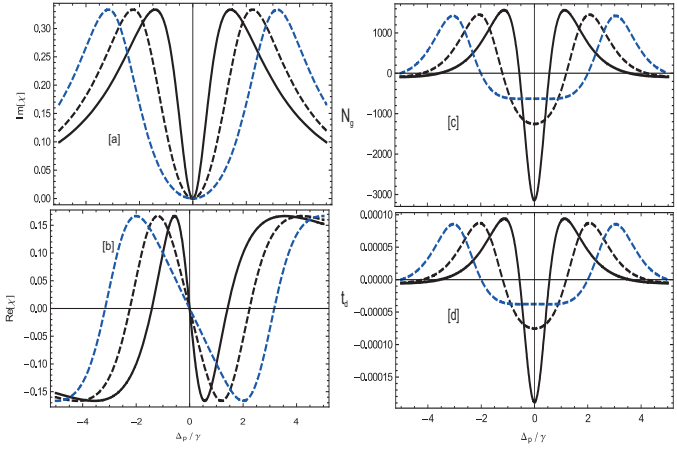}
\caption{ Absorption dispersion, group index and group delay time
against $\frac{\Delta_p}{\protect
\gamma}$ such that $\protect\gamma%
=1MHz$, $\gamma_{1}=\gamma_{1}= \gamma_{3}=2\protect\gamma$,
$\protect\omega_{ac}=10^3 \protect\gamma$, $
\protect\Delta_1=0\protect\gamma$,
$\protect\Delta_2=0\protect\gamma$, $ \Omega_1=2\protect\gamma$,
$\Omega_2=2\protect\gamma$ (solid black), $4\protect\gamma$(black
dashed), $6\protect\gamma$(blue dashed)} \label{figure1}
\end{figure}

\begin{figure}[t]
\centering
\includegraphics[width=3.5in]{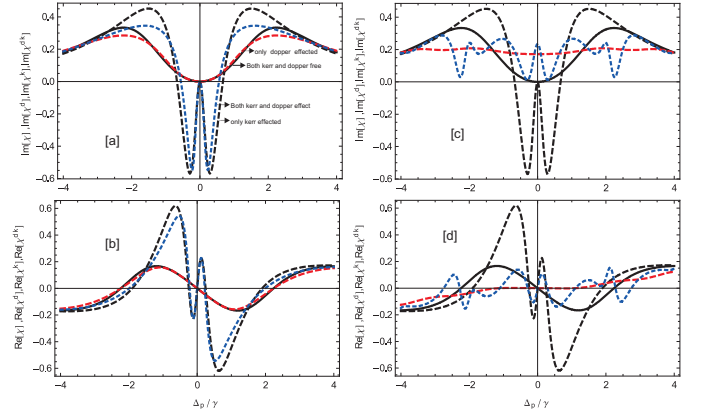}
\caption{ Absorption and dispersion vers $\frac{\Delta_p}{\protect
\gamma}$ such that $\protect\gamma%
=1MHz$, $\gamma_{1}=\gamma_{1}= \gamma_{3}=2\protect\gamma$,
$\protect\omega_{24}=10^3 \protect\gamma$, $
\protect\Delta_1=0\protect\gamma$,
$\protect\Delta_2=0\protect\gamma$, $ \Omega_1=4\protect\gamma$,
$\Omega_2=2\protect\gamma$, $V_D=2\gamma$, [a,b]$\alpha_{i=1,2}=1$
[c,d]$\alpha_{i=1,2}=-1$. Niether Doppler no kerr effect black
solid line. Doppler effect but no kerr effect black dashed line.
Kerr effect but no Doppler effect red dashed line. Both Doppler
and kerr effect dashed blue line.} \label{figure1}
\end{figure}
\begin{figure}[t]
\centering
\includegraphics[width=3.5in]{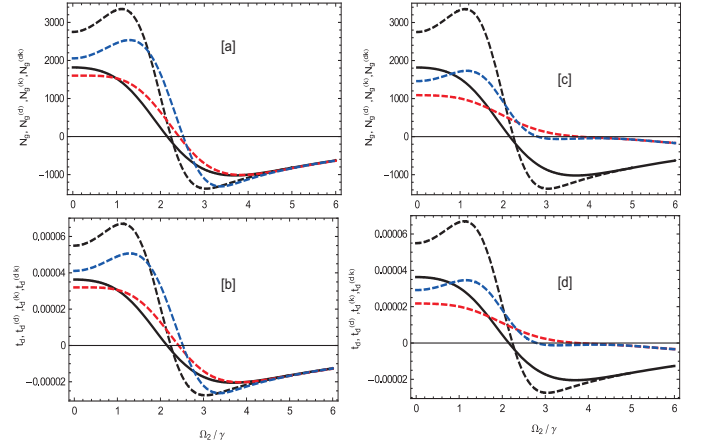}
\caption{ Group index and group delay/advance time vers
$\frac{\Omega_2}{\protect
\gamma}$ such that $\protect\gamma%
=1MHz$, $\gamma_{1}=\gamma_{1}= \gamma_{3}=2\protect\gamma$,
$\protect\omega_{24}=10^3 \protect\gamma$, $
\protect\Delta_1=0\protect\gamma$,
$\protect\Delta_2=0\protect\gamma$, $ \Omega_1=2\protect\gamma$,
$\Delta_p=0.6\protect\gamma$, $V_D=2\gamma$,
[a,b]$\alpha_{i=1,2}=1$ [c,d]$\alpha_{i=1,2}=-1$. Niether Doppler
no kerr effect black solid line. Doppler effect but no kerr effect
black dashed line. Kerr effect but no Doppler effect red dashed
line. Both Doppler and kerr effect dashed blue line.}
\label{figure1}
\end{figure}
\begin{figure}[t]
\centering
\includegraphics[width=3.5in]{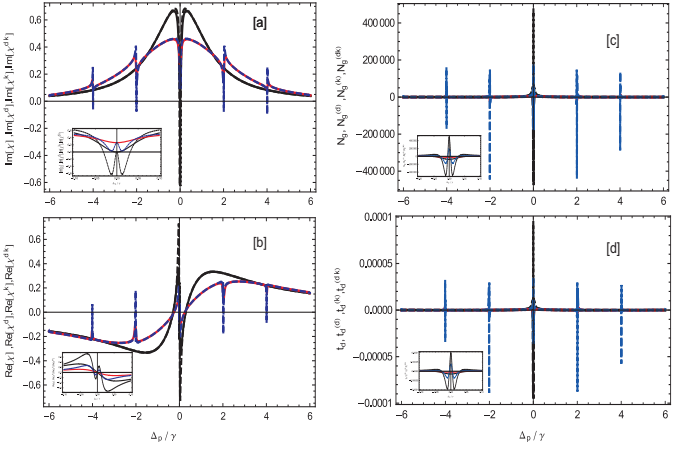}
\caption{Absorption, Dispersion, Group index and group
delay/advance time vers $\frac{\Delta_p}{\protect
\gamma}$, such that $\protect\gamma%
=1MHz$, $\gamma_{1}=\gamma_{1}= \gamma_{3}=1\protect\gamma$,
$\protect\omega_{24}=10^3 \protect\gamma$, $
\protect\Delta_1=0\protect\gamma$,
$\protect\Delta_2=0\protect\gamma$, $ \Omega_1=0.5\protect\gamma$,
$\Omega_2=0.3\protect\gamma$, $V_D=1.5\gamma$
[a,b,c,d]$\alpha_{i=1,2}=-1$. Niether Doppler no kerr effect black
solid line. Doppler effect but no kerr effect black dashed line.
Kerr effect but no Doppler effect red dashed line. Both Doppler
and kerr effect dashed blue line.} \label{figure1}
\end{figure}

\begin{figure}[t]
\centering
\includegraphics[width=3.5in]{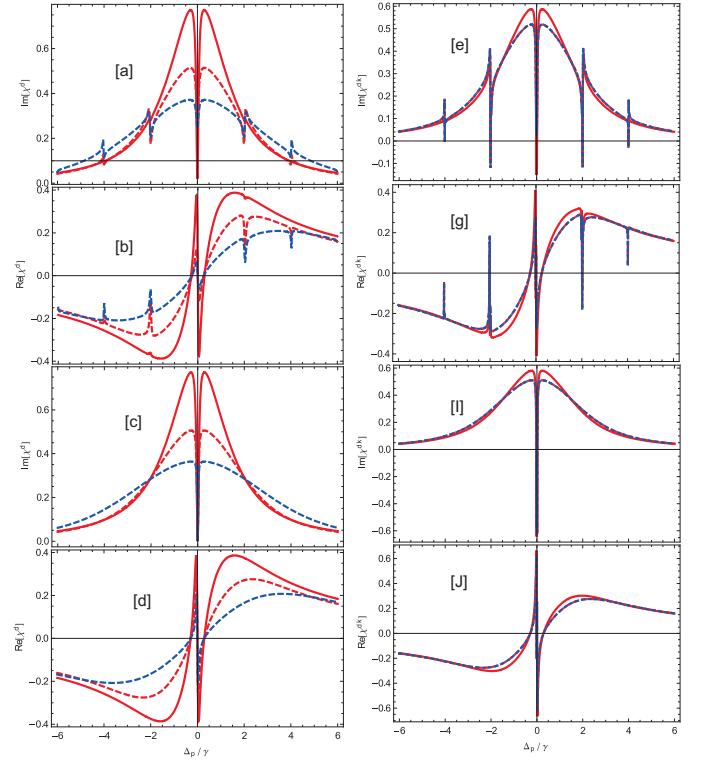}
\caption{Absorption and dispersion  vers $\frac{\Delta_p}{\protect
\gamma}$ such that $\protect\gamma%
=1MHz$, $\gamma_{1}=\gamma_{1}= \gamma_{3}=1\protect\gamma$,
$\protect\omega_{24}=10^3 \protect\gamma$, $
\protect\Delta_1=0\protect\gamma$,
$\protect\Delta_2=0\protect\gamma$, $ \Omega_1=0.5\protect\gamma$,
$\Omega_2=0.3\protect\gamma$, $V_D=1.5\gamma$
[a,b,e,f]$\alpha_{i=1,2}=-1$[c,d,I,j]$\alpha_{i=1,2}=1$,$V_D=0.5\gamma$
(solid red), $V_D=1.5\gamma$(Dashed red), $V_D=3\gamma$(blue
dashed).} \label{figure1}
\end{figure}
We explain our main results in the Eqs.(13,15,17,18,19,20,21,23)
for absorption, dispersion, group index, time delay and pulse
shape distortion, when there is no kerr effect neither Doppler
broadening effect in the system. We also present our results for
Absorption, Dispersion, Group index and for the system when there
is its maximum kerr and broadening effects. When we put $v=0$, in
the eq13, we obtain the optical results of double lamda
configuration for cold atomic system. Further if we turn off one
of the control field in our system [$\Omega_1=0$, or
$\Omega_2=0$], the optical behavior of our system concise with the
famous \textit{A-type} atomic configuration. Next, we focus on the
main results of our atomic system and are committed (1) to discuss
the subluminal and superluminal behavior of the light pulse
propagating through their associated dispersive regions, (2) to
discuss the incoherent Doppler broadening effect in the system,
(3) to discuss the kerr effect in the system. The kerr effect is
coherent effect and significant contribution to optical
properties. Experiment can be easily adjusted on kerr effect. The
Doppler broadening effect is temperature dependent incoherent and
experiment is not so easy as compare to kerr effect. In Fig2 , the
plots are traced for the system, when there is no kerr effect
neither Doppler broadened.
 The absorption spectrum show a typical transparency for the propagating probe field at resonance
 $\Delta_p=0$, as shown in Fig. 2a. The slope of dispersion in transparency window is anomalous shown in [Fig.
 2b]. The group index in the anomalous region is negative. The
 value of group index and group advance time at the parameters
 $\gamma_1=\gamma_2=\gamma_3=2\gamma$ and
 $\Omega_1=\Omega_2=2\gamma$ are $-3000$ and $-200\mu s$ in the
 medium as shown in [Fig.2(c,d)]. When the intensity of the control field $\Omega_2$
are increase from $2\gamma$ to
  $4\gamma,6\gamma$, the
 transparency width is increase and the negative group index and
 time advancement are degraded due to the less anomalous dispersion
 see in Fig[a,b,c,d] black dashed and blue dashed lines.
 Fig3[a,b,c,d] show absorption and dispersion spectrums for Doppler and kerr free system by solid black
 line. The Doppler free kerr effected system by dashed black line. The Doppler effected but kerr free system by dashed
red line. The Doppler as well as kerr effected system by dashed
blue line. The joint effect of kerr nonlinearity and Doppler
broadened introduced in the system, show significant special
profiles of absorption and dispersion. The absorption in the
resonance point $\Delta_p=0$, in all the cases are approach to
zero. At this point the light are totally transmitted.
 Around the resonance point the absorption is a function of probe
 detuning $\Delta_p$, and different spectral profiles with the kerr
 and Doppler broadened effect. Closed to the resonance point at
 $\Delta_p=\pm0.25$, there are large symmetric negative absorption called Electromagnetically
 induced transparency Amplification with kerr
 nonlinearity, if both control and kerr field co-propagates ($\alpha_i=1$), with the probe field
 as shown in Fig3[a]. The slope of dispersion at resonance point $\Delta_p=0$,
 is anomalous in the kerr free system (cold and hot atomic system with no kerr effect).
  When the kerr effect is switch in the system the
 slope of dispersion is reversed (normal) at resonance $\Delta_p=0$, both in Doppler free and Doppler effect
 system. Near the resonance point at $\Delta_p=\pm2.5$, the slope
 of dispersion are anomalous in all the cases but steep anomalous with kerr effect
 Fig3[b]. The normal dispersion show slow light propagation, while
 anomalous dispersion describe fast light propagation. When the
 control and kerr field courter propagate ($\alpha_1=-1$), to the probe field in
 the cell, the the negative absorption is reduced and vanished at
 a certain intensity of the kerr field. The absorption are
 increase with the Doppler effect and small symmetric
lamb dip are appears on both sides of the central absorption peak.
The central absorption peak and Lamb dips are enhances with the
strength of kerr effect, see in Fig3[c]. The dispersion slopes are
normal at the resonance in the presence of kerr effect, while
anomalous without kerr effect for both cold and hot medium. To
studies the important and detail physic of normal and anomalous
dispersion and their variation with Doppler as well as kerr effect
we plots the group index and group delay times near the resonance
point against the control field $\Omega_2$, as shown in
Fig4[a,b,c,d], when the control and kerr field co-propagating
($\alpha_{i=1,2}=1$), to the probe field. The group index as well
as time delay/advance are the function of control field keeping
the probe detuning $\Delta_p=0.5\gamma$. At low intensity, bellow
 $\Omega_2=2\gamma$, the group index and
time delay are positive. The values of group index for all the
four cases at $\Omega_2=1.5\gamma$ are written here $N_g=986.90$
,$N^k_g=2945.95$, $N^d_g=1244.91$, $N^{d,k}_g=2472.8$ as shown in
Fig4[a]. The corresponding group delay time are $t_d=19\mu s$,
$t^k_d=58\mu s$, $t^d_d=24\mu s$, $t^{d,k}_d=49\mu s$, see in the
Fig4[b]. At the positive group index, group velocity is
$v_g=c/N_g$, is positive and varies with the group index. The
pulses of all the four cases are different group velocities of
[$c/986.90$, $c/2945.95$, $c/1244.91$, $c/2472.8$]. These results
have large potential application in telecommunication systems. It
is important to the preservation of two light pulses. If both
pulses are arrive at the same time to the detectors, they will
accumulate only one of them, and the information of the other will
be lost, which slow down the over all flow of information.
Activating the slow light medium, one of the pulse is delay from
the other, the importation of both the pulses are preserved, while
the flow of information is speed up. To discus the advantage of
negative group index and advance time observe the same two
Fig4[a,b]. Above the intensity of the control field
$\Omega_2=2\gamma$, the group index and time delay are
negative(advance time) values. The values of group index for all
the four cases at $\Omega_2=3\gamma$ are $N_g= -862.84$ ,$N^k_g=
-1367.93$, $N^d_g=-703.013$, $N^{d,k}_g=-1104.11$, see in Fig4[a].
The corresponding group advance times are $t_{ad}=-17.27\mu s$,
$t^k_d=-27.37\mu s$, $t^d_d=-14.80\mu s$, $t^{d,k}_d=-22.10\mu s$
see in Fig4[b]. The negative time delay (advance time) are very
important for spacial modes images. The degree of spacial modes
images are quantified and increase in qualities, when the advance
time is increase. Advance time are also potential application in
spacial and temporal cloaking devices, therefore our results will
be easily adjusted to improve the current technology of
telecommunication systems and images as well as cloaking. At high
sufficient value of control field the group index and group delay
are then saturated. Fig5 show variation of absorption, dispersion,
group index and group delay/advancement with probe detuning. When
the rabbi ($\Omega_{i=1,2}$), frequencies of control and kerr
field are smaller then decay rates ($\gamma_{i=1,2,3}$), while the
kerr and control field counter-propagate ($\alpha_i=-1$) to the
probe field. The Dark lines are appears on the spectrum due to
fano type quantum interference. These dark lines are more dominant
with doppler effect. These different dark lines arise due to
quantum interference effect among the excitation probability
amplitudes of the probe field. The probe field follows different
paths in the presence of Doppler broadened, by the dressed states
of the excited real energy level created by the atom-field
interaction of the system. The Lorentzian line shape of the
spectra split into different components, which depend on the
numbers of dark lines and its sum of the FWHM, obey the
Weiskopf-Wigner theory, see in Fig5[a,b]. The group index close to
the resonance region at $\delta_p=0.01$, are negative for all the
cases. The values of group index for all the four cases at this
nearest resonance point of the probe detuning, are $N_g=-67580$,
$N^k_g=-459616$, $N^d_g=-25046$, $N^{d,k}_g=-172503$ as shown in
Fig5[c]. The group velocity($c/N_g$) at this region are
$v_g=-c/67580$, $v^k_g=-c/459616$, $v^d_g=-c/25046$,
$v^{k,d}_g=-c/172503$. It mean that the kerr effected medium
without Doppler broadened (cold kerr atomic medium) is more
superluminal. The significant advance times of the pulse for the
four cases are $t_{ad}=-13\mu s$, $t^k_d=-91\mu s$, $t^d_d=-5\mu
s$, $t^{d,k}_d=-34\mu s$, see in Fig5[d]. These results are
applicable to increase the quality of spacial modes images.
Further the results are potential application for the temporal
cloaking devices to increase the time gaps. In Fig6, the graphs
are traced for absorption and dispersion versus probe detuning.
The absorption and dispersion show intrusting behavior with the
Doppler and kerr effect. The  topical Lamb dip at the resonance
point are more dominant with small value of the Doppler width.
When the Doppler width is increase the lamb dip is reduced. The
superluminal behavior are degraded with the decrease of Lamb dip.
The Dark line are appear on the spectrum when the coherent fields
are counter propagate to the probe field. These dark lines are
enhanced value with the kerr nonlinearity. The subluminal and
superluminal behavior are enhance, when the hight of dark lines
are increase. Fig7 show Gaussian pulse shape at the input and out
put of the medium. All the pulses are the same shape, when the
coherent fields counter propagate to the probe (a) and
co-propagate to the probe but slightly broaden. Where $\tau_0$ is
input pulse width. In signal processing, optical memories, data
synchronization, optical buffers, cloaking devices, spacial modes
images required slow and fast light pulse delay/advance $t_d$,
times. For significant process the delay/advance $t_d$, time need
to be very large as compare to the pulse with $\tau_0$. Other
requirement is that the pulse not be distorted. These two
condition oppose each other. Large delay occur with great
distortion. Hoverer the distortion can be minimized. Therefore the
system is sufficient distortion, which does not effect the
subluminal and superluminal propagation. In conclusion, we
proposed a double lambda 4-level atomic system driven by two
coherent control fields, and a probe field. The proposed scheme
displays interesting results of subluminal and superluminal light
with the jointly effect of kerr nonlinearity and Doppler
broadening. These results are various advantages in the
telecommunication process as well as spacial modes image and
temporal cloaking devices.

\end{document}